\documentclass[mathleft
]{an}
\usepackage{graphics}
\usepackage{graphicx}
\usepackage{amsmath}
\usepackage{times}
\usepackage[english]{babel}
\begin{document}

\Pagespan{789}{}
\Yearpublication{2006}%
\Yearsubmission{2005}%
\Month{11}%
\Volume{999}%
\Issue{88}%

\title{Control of chaos in the vicinity of the Earth--Moon {$L_5$} Lagrangian point to keep a spacecraft in orbit}
\author{J. Sl\'{i}z\inst{1}\thanks{Corresponding author:
  \email{judit.sliz@gmail.com}\newline}
\and \'A. S\"uli\inst{1}
\and T. Kov\'acs\inst{2}
}
\titlerunning{Control of chaos}
\authorrunning{J. Sl\'{i}z}
\institute{
Department of Astronomy, E\"otv\"os University, P\'azm\'any P\'eter s\'et\'any 1/A, 1117 Budapest, Hungary\\
Konkoly Observatory, Research Centre for Astronomy and Earth Sciences, Hungarian Academy of Sciences, PO Box 67, H-1525 Budapest, Hungary}

\received{30 May 2005}
\accepted{11 Nov 2005}
\publonline{later}


\abstract{%
The concept of Space Manifold Dynamics is a new method of space research. We have applied it along with the basic idea of the method of Ott, Grebogi and York (OGY method) to stabilize the motion of a spacecraft around the triangular Lagrange point {$L_5$} of the Earth--Moon system. We have determined the escape rate of the trajectories in the general three- and four-body problem and estimated the average lifetime of the particles. Integrating the two models we mapped in detail the phase space around the{$L_5$} point of the Earth--Moon system. Using the phase space portrait our next goal was to apply a modified OGY method to keep a spacecraft close to the vicinity of {$L_5$}. We modified the equation of motions with the addition of a time dependent force to the motion of the spacecraft. In our orbit--keeping procedure there are three free parameters: ({\it i}) the magnitude of the thrust, ({\it ii}) the start time and ({\it iii}) the length of the control. Based on our numerical experiments we were able to determine possible values for these parameters and successfully apply a control phase to a spacecraft to keep it on orbit around {$L_5$}.
}
\maketitle

\section{Introduction}

The concept of Space Manifold Dynamics (SMD) encompasses the various applications of methods to analyze and design spacecraft missions, ranging from the exploitation of libration orbits around the collinear Lagrangian points (denoted by ${L_1, L_2, L_3)}$ to the design of optimal station--keeping and eclipse avoidance manoeuvres or the determination of low energy lunar and interplanetary transfers (\cite{Perozzi2010}). 
\par
The classical calculations of interplanetary trajectories (e.g. Apollo lunar missions, the Voyagers grand tour, or the New Horizons mission) are based patched-conics transfers and using the "gravity assist" principle exploiting the gravitational force of planets. The demand for new and unusual kinds of orbits to meet their scientific goals made it necessary to look for a new approach which uses the dynamics of manifolds. The available tools allows researchers to study and analyse the natural dynamics of the problem and provides a way to, for instance, design new "low-energy" orbits, cheap station keeping procedures and eclipse avoidance strategies. These tools are based on the examination of the stable and unstable manifolds of the system.
\par
The International Sun/Earth Explorer 3 (ISEE-3, launched on August 12, 1978) was the first artificial object placed at ${L_1}$, proving that such a suspension between gravitational fields was possible. The Solar and Heliospheric Observatory (SOHO, launched on December 2, 1995) was the successor of the ISEE-3 (\cite{Dunham2003}). The SMD approach was first applied to the SOHO program (\cite{Gomez2001}). SOHO was in {$L_1$} and explored the Sun--Earth interactions, first examining the phenomenon known as "space weather" now. After the success of these missions the scientific interest increased towards the benefits of Lagrangian points and the new method was used to design more and more ambitious missions that were later realized in the MAP, Genesis, and Herschel--Plank missions. Genesis was the first spacecraft whose orbit has been entirely designed with the new method (\cite{Howell1998}).
\par
The Lagrangian points have obvious benefits for space missions. For example, the ${L_1}$ point of the Earth--Sun system affords an uninterrupted view of the Sun and is currently home to the SOHO. The ${L_2}$ point of the Earth--Sun system was the home to the WMAP spacecraft, current home of Planck, and future home of the James Webb Space Telescope. ${L_2}$ is ideal for astronomy because a spacecraft is close enough to readily communicate with Earth, can keep Sun, Earth and Moon behind the spacecraft for solar power and (with appropriate shielding) provides a clear view of deep space for our telescopes. The ${L_1}$ and ${L_2}$ points are unstable on a time scale of approximately 23 days, which requires satellites orbiting these positions to undergo regular course and attitude corrections. It is unlikely to find any use for the ${L_3}$ point since it remains hidden behind the Sun at all times. The idea of a hidden "Planet-X" at the ${L_3}$ has been a popular topic in science fiction writing. The instability of Planet X's orbit (on a time scale of 150 years) didn't stop Hollywood from shooting movies like The Man from Planet X.
\par
The ${L_4}$ and ${L_5}$ points are home to stable orbits so long as the mass ratio between the two large masses  exceeds 24.96 {(in our case between the Earth and the Moon it is 81}) (\cite{Murray1999}). This condition is satisfied for both the Earth--Sun and Earth--Moon systems, and for many other pairs of bodies in the solar system. In 2010 NASA's WISE telescope finally confirmed the first Trojan asteroid (2010 TK7) around Earth's leading Lagrange point. 
\par
The stability property of ${L_4}$ and ${L_5}$ offers an excellent opportunity for spacecrafts to maintain their orbits with minimal fuel consumption. These points could be excellent locations to place space telescopes for astronomical observations or a space station (\cite{Schutz1977}). In addition, there is the renewed interest of major space agencies for Lagrangian point colonization. Furthermore, \textit{\cite{Fillipi1978}} has made a review of the ideas of \textit{\cite{Neill1974}} about building space colonies at the $f{L_4}$ and ${L_5}$ positions. These space stations could be used as a waypoint for travel to and from the cislunar space. Currently, there are no probes orbiting ${L_4}$ or ${L_5}$ points for any celestial couple. \textit{\cite{Salazar2012}} proposed an alternative transfer strategy to send spacecraft to stable orbits around the Lagrangian equilibrium points ${L_4}$ and ${L_5}$ based on trajectories derived from the periodic orbits around ${L_5}$. In the present work or main goal is to provide a method for keeping a station around the orbit of ${L_5}$.
\par
In Section 2 we describe our models and intial conditions. In Section 3 we present the results of the simulations as phase portraits in the vicinity of ${L_5}$ and give the escape rate and the associated average lifetime of the test particles in the different models. The orbit--keeping procedure is demonstrated with the use of a test case and described in detail in Section 4. Finally our conclusions are given in Section 5.

\section{Model and initial conditions}

\begin{figure}
\includegraphics[width=\linewidth]{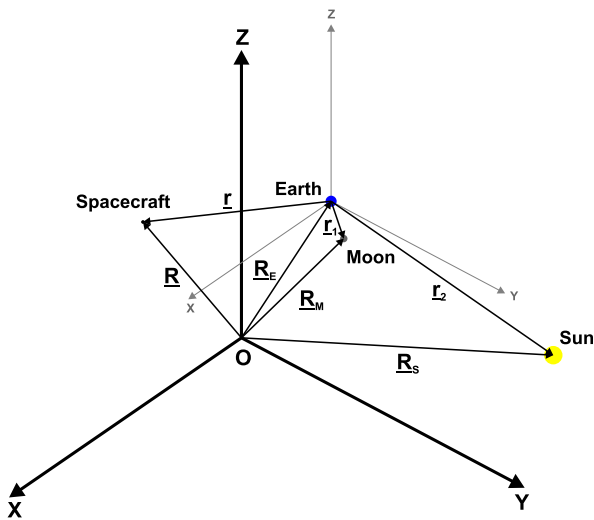}
\caption{The model in the inertial reference frame $OXYZ$: ${\vec{R}},\,{\vec{R_E}},\,{\vec{R_M}},\, {\vec{R_S}}$ denotes the position vectors of the spacecraft, Earth, Moon and the Sun respectively. The vectors ${\vec{r}}, {\vec{r_1}}$ and ${\vec{r}_2}$ denote the geocentric positions of the spacecraft,  Moon and the Sun respectively.}
\label{Fig1}
\end{figure}

In the present study two models were used: the (restricted) three body-problem (R3BP) (\cite{Szebehely1967}) and (restricted) four body-problems (R4BP). In our simulations one of the bodies, the spacecraft has negligible mass, but we have kept its mass in the equations of motion in order to be able to compute the fuel consumption (hence the restricted adjective is in parentheses). The R3BP models the motion of a particle under the gravitational attraction of two larger bodies (also called primaries), where the primaries are point masses that revolve in elliptic orbits around their common centre of mass. In our case the primaries were the Earth and the Moon. In order to take into account of the effect of the Sun the model was extended to the R4BP. The equations of motion are valid in 3D and were transformed to the geocentric reference frame. The model is depicted in Fig. \ref{Fig1}, where ${\vec{R}},\,{\vec{R_E}},\,{\vec{R_M}},\, {\vec{R_S}}$ denotes the position vectors of the spacecraft, Earth, Moon and the Sun in an inertial reference frame. The relative vectors ${\vec{r}} = (x,y,z)$ ${\vec{r}_1}$ and ${\vec{r}_2}$ denote the geocentric position vector of the spacecraft, Moon and the Sun respectively.
\par
\begin{figure}
\includegraphics[width=\linewidth]{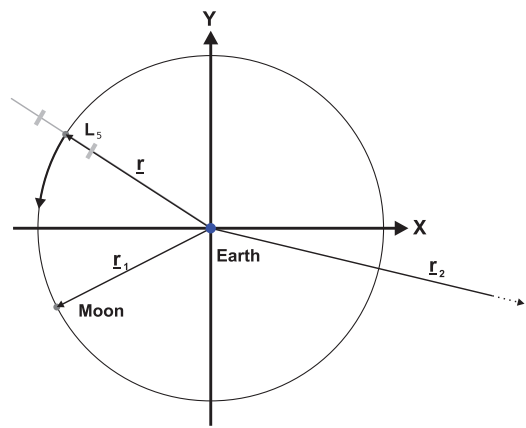}
\caption{Initial conditions: top view of the geocentric system. The initial positions were chosen from between the light gray bars located on the line connecting the Earth and the ${L_5}$ point. The notion are the same as in Fig. \ref{Fig1}.}
\label{Fig2}
\end{figure}

The first three Lagrangian equilibrium points, ${L_1, L_2}$  and ${ L_3}$ are aligned with the two larger bodies. On the other hand, ${L_4}$ and ${L_5}$ points form equilateral triangles with the two masses on the plane of orbit whose common base is the line between their center of mass and ${L_4}$ leads the orbit of the smaller mass and ${L_1}$ follows (Fig. \ref{Fig2}).
\par
Of particular interest is the stability of the Lagrangian points: the three collinear stationary points are always unstable, whereas the stability of ${L_4}$ and ${L_5}$ points depends on the mass ratio between the two larger bodies  (\cite{Danby 1964}) and eccentricity. In the case of the important celestial couples like Earth-–Moon or Sun-–Earth ${L_4}$ and ${L_5}$ points are stable. This property makes the fuel required for a spacecraft to maintain its relative position there to be almost zero.
\par
The first goal was to map the phase space around the ${L_5}$ point in the RTPB of the Earth--Moon system. To explore the stable and unstable regions of the motion of the spacecraft in the R3BP model, the equations of motion have to be integrated for a large set of initial conditions. For the given initial condition of the Moon (see Table \ref{Table1}), we integrated orbits of the test particle placing it at different points as starting points along the line connecting ${L_5}$ and the Earth as demonstrated in Fig. \ref{Fig2}. The initial velocity was chosen such that the angular velocity of the test particle was equal to that of the ${L_5}$ point. The time of integration was 1300 days (for several test cases it was extended to 144 000 days). The determination of the length of the integration time is explained in Section 3.1. This initial condition generation procedure was applied also in the simulations of the R4BP (see Fig. \ref{Fig3}).
\par
The major goal of the present study is to present a procedure to keep a probe on orbit around the EML5\footnote{The five Sun--Earth Lagrangian points are called SEL1--SEL5, and similarly those of the Earth--Moon system EML1--EML5, etc.} point of the Earth--Moon system. The initial condition for the satellite is given in Table \ref{Table1}.
\par
We applied a variable step-size Runge-Kutta method (\cite{Dormand1978}). The initial conditions of the major bodies are from the JPL database with epoch = 2455999.5, in geocentric ecliptic Cartesian coordinate system, see Table \ref{Table1}.
\begin{table*}[h!]
\begin{tabular}{llll}
          & Moon                & Sun                  & Spacecraft          \\
\hline
$x$       & -0.5166166629896271 & 368.66644026568872   & -0.9141820107443692 \\
$y$       & -0.7573376905382426 & -47.60083686893512   &  0.0687343088982397 \\
$z$       & -0.0172345994799509 &   0.0004280643569276 &  0                  \\
\hline
$\dot{x}$ &  0.185382724767311  &   0.9306857451293241 & -0.0252733218674244 \\
$\dot{y}$ & -0.1362138844144604 &   6.4039891664344024 & -0.2286530912785001 \\
$\dot{z}$ &  0.0197096442582322 &   0.0000507123227653 &  0                  \\
\hline
\end{tabular}
\caption{The initial condition of the bodies taken from the JPL. The values are valid for JD = 2455999.5. The initial condition for the spacecraft is in the close vicinity of ${L_5}$.}
\label{Table1}
\end{table*}

\section{Results}

\begin{figure}
\includegraphics[width=\linewidth]{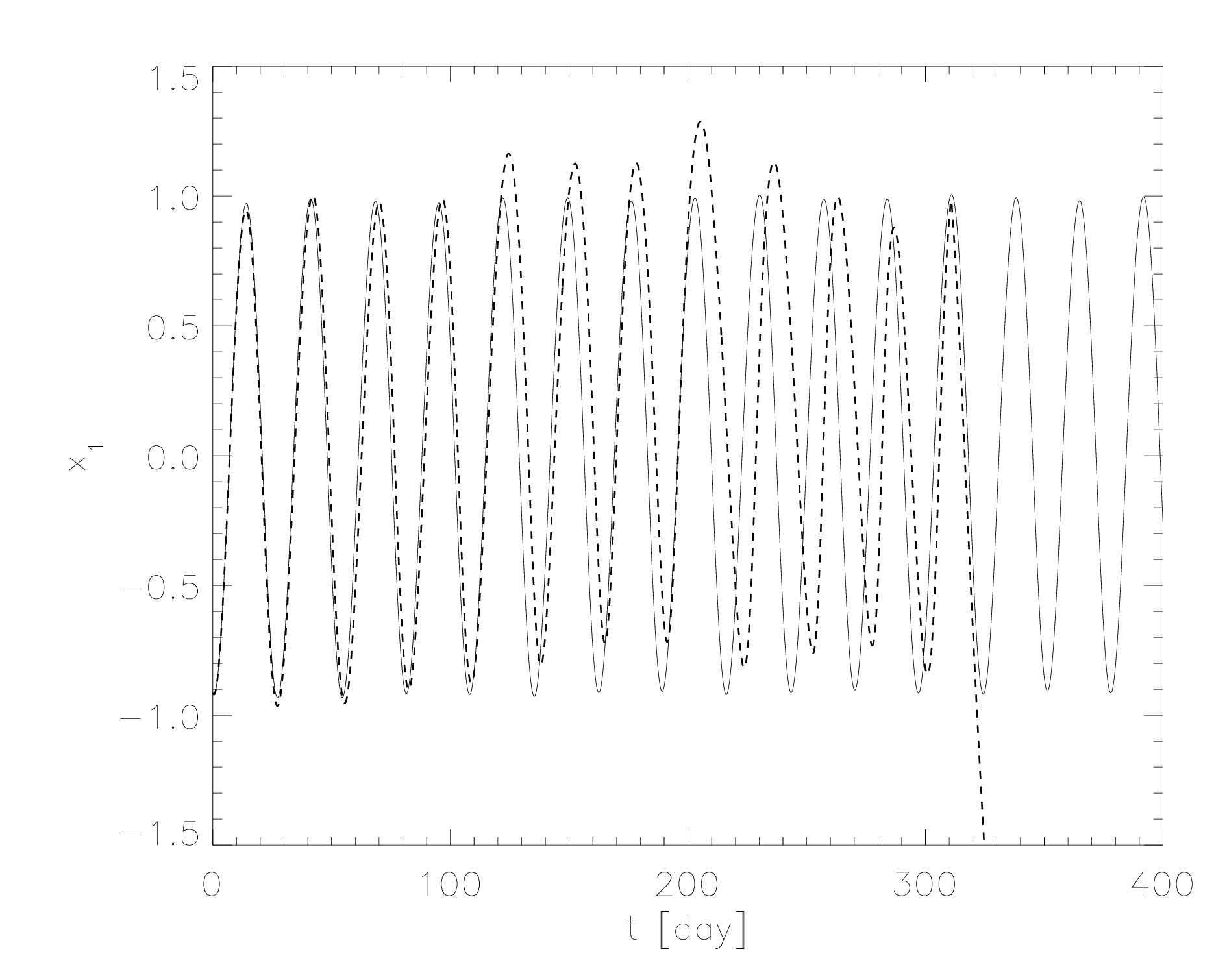}
\caption{The $x_1(t)$ of the test particle in the R3BP (continuous line) and in the R4BP (dashed line). The test particle was started from the vicinity of the ${L_5}$ point of the planar R3BP Earth--Moon system.} 
\label{Fig3}
\end{figure}

The elliptic restricted three-body problem is often used in dynamical investigations of planetary systems or to design transfer strategy to send spacecrafts to stable orbits around one of the Lagrangian points. However, in real missions other perturbations must be taken into account. In the present study we include the most relevant force, which is the Sun's gravitational attraction. Fig. \ref{Fig3} clearly demonstrates the crucial role the Sun plays in the orbital evolution of a probe: in the R3BP the probe moves on a regular orbit, whereas in the R4BP its trajectory is chaotic and the spacecraft escapes in less than 1 year.
\par
First we mapped a larger region of the phase space around ${L_5}$ in different models. The aim was to acquire a general picture of the structure of the phase space: where are the stable and chaotic regions and what are their sizes and shapes. Also we were interested in the escape rates and in the average lifetime of the test particles. This knowledge is essential to design the orbit and the orbit--keeping strategy of a probe.

\subsection{Transient chaos and escape rates in the vicinity of ${L_5}$}

In transient chaos, typical trajectories, i.e. trajectories initiated from random initial conditions, escapes any neighborhood of the non attracting chaotic set (\cite{Kovacs2009}). To define the escape rate one distributes a large number of initial points $N_0$ at time $t_0$ in a phase space region $R$ and then follows the trajectories of these points. A quantity measuring how quickly particles leave $R$ is the escape rate $\kappa$:
\begin{equation}
 N(t) = N_0e^{-\kappa (t-t0)},
\end{equation}
where $N(t)$ is the number of particles which are still part of the initial region $R$. A small value of $\kappa$ implies weak "repulsion" of typical trajectories by the non attracting chaotic set. The escape rate is a property solely of the non attracting chaotic set.
\par
\begin{figure*}
\includegraphics[width=80mm]{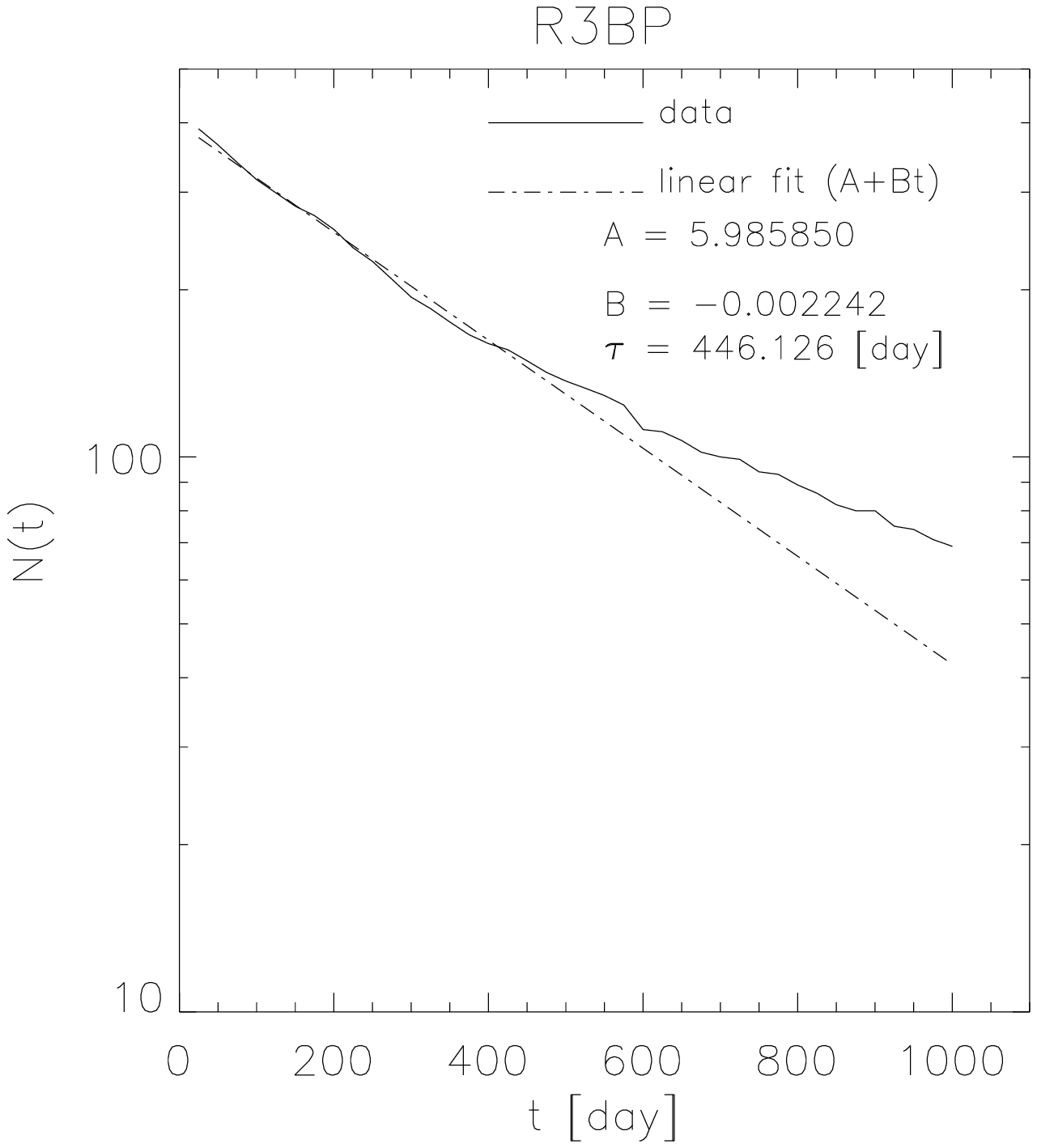}
\includegraphics[width=80mm]{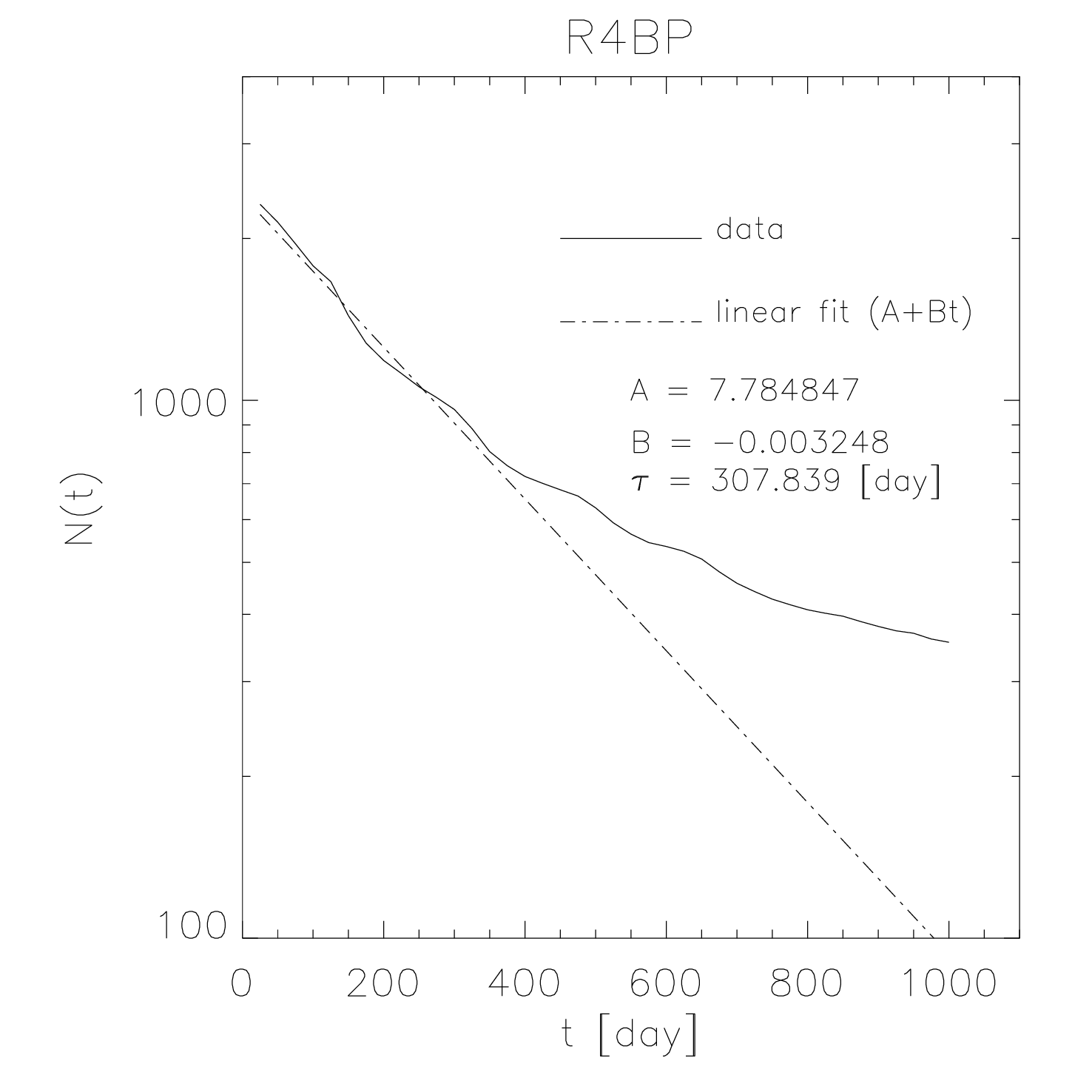}
\caption{The escape rates for the models R3BP (left panel) and R4BP (right panel). The continuous curve is the measured data, the dashed line is the fitted line $N(t) = A + Bt$. The parameters of the line $A$ and $B$ as well as $\tau = -1/B$ are displayed on each panel (see also Table \ref{Table2}).}
\label{Fig4}
\end{figure*}

To find $\kappa$ for the R3BP using the initial conditions described in the previous section the space region $R$ was determined. Using a regular grid covering $R$ the trajectories of the test particles were followed for 1 000 days both in the R3BP and R4BP.   The number of test particles we used for the R3BP is 400 and for the R4BP is 10 000.
Our main model was the R4BP, we investigated it first and we used more initial points.  After receiving the
result, (Fig. 4b), we saw that for the R3BP will be enough less initial points to get the escape rate, because it is defined only by the slope of the line.
The results are shown in Fig. \ref{Fig4} left and right panel, respectively. The average lifetime $\tau$ is estimated as $\tau \approx 1/\kappa$. We have fitted paired data $\{t_i, {\mathrm ln}(N(t_i)\}$ to the linear model, $y = A + Bx$, by minimizing the chi-square error statistic. The slope of the line is the parameter $\kappa$ and its reciprocal is the average lifetime. The average lifetime of the test particles are approximately 440 days ($\approx$ 16 revolution of the Moon) and 310 days ($\approx$ 11 revolution of the Moon) in R3BP and R4BP, respectively. The  later is significantly shorter than that in R3BP. The parameters of the fitted line and $\tau$ are summarized in Table \ref{Table2}.
\begin{table}
\begin{tabular}{llll}
\hline
Model & $A$      & $B$          & $\tau$ [day ] \\
\hline
R3BP  &  5.9858  & -0.00224152  & 446.126 \\
R4BP  &  7.7848  & -0.00324845  & 307.839 \\
\hline
\end{tabular}
\caption{Results of the linear fit to the models R3BP and R4BP. In the first column the model in the second the constant $A$ and the third column displays the slope of the fitted linear model. The escape times are given in the last column.}
\label{Table2}
\end{table}

\par
\begin{figure*}
\includegraphics[width=\linewidth]{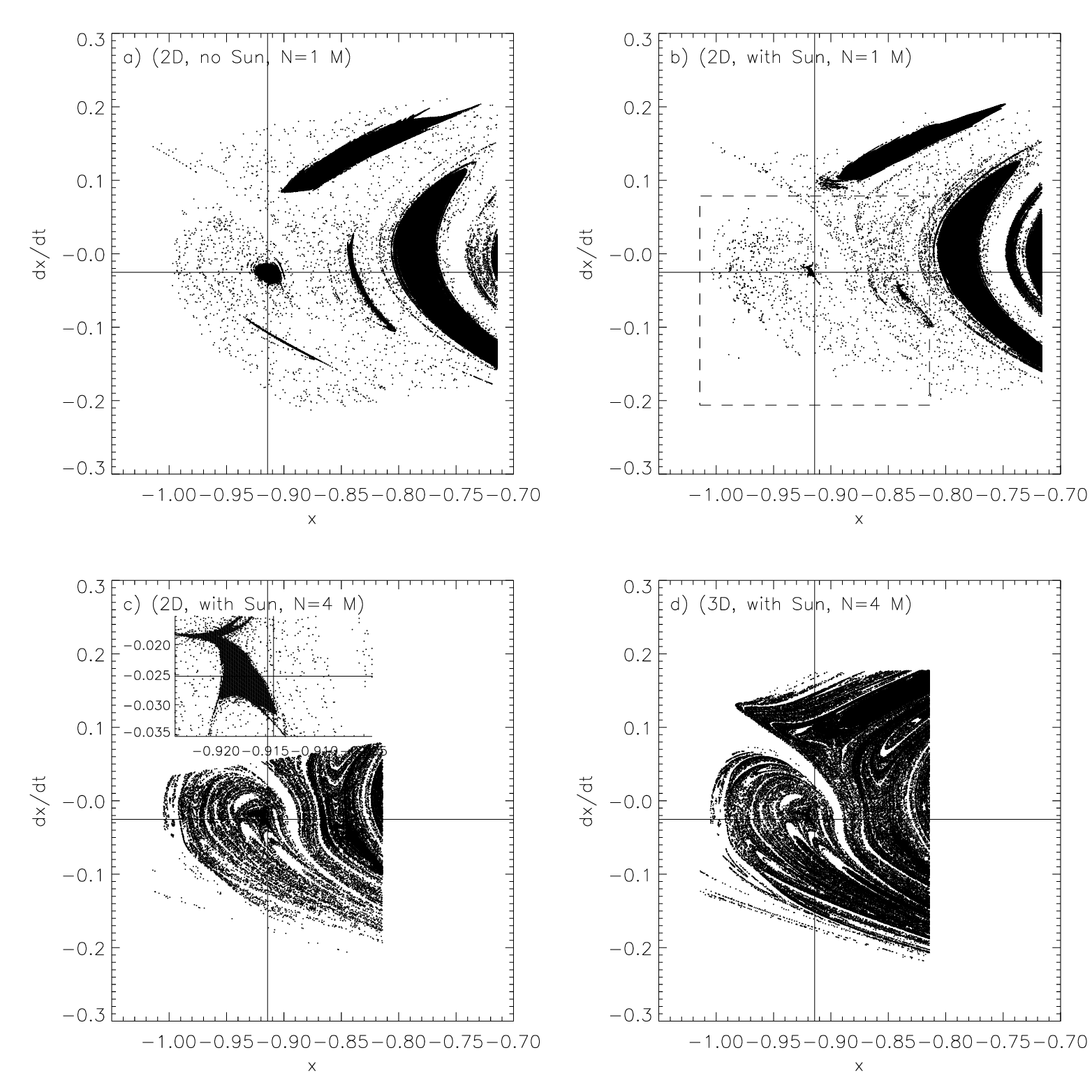}
\caption{The $x - \dot{x}$ phase space. The points corresponds to orbits which did not leave the $x \in [-2,2]$ and $\dot{x} \in [-2,2]$ region after 1300 days. The legend shows the dimension of the simulation (2D or 3D), whether the Sun was included or not, (R3BP or R4BP) and the number of initial points ($N$). The position of the ${L_5}$ point is shown by a large cross. Panel a) shows the result of the planar R3BP; b) shows the result of the planar R4BP; c) shows a smaller region around ${L_5}$ in the planar R4BP, while d) shows the result of the spatial R4BP for 4 million test particles.} 
\label{Fig5}
\end{figure*}

According to the results obtained for the average lifetime the simulation time was set as usual the 4--6$\times \tau$ (\cite{Tel_2006}; \cite{Lai2011}), i.e. in all our simulations the test particles were integrated for $T=1300 (\approx 5\tau)$ days. The results are shown in Fig. \ref{Fig5}. In the figures the horizontal axis is the test particle's initial $x$ coordinate in units of the average distance $r_0 = 400 000$ km of the Earth--Moon. The vertical axis is the test particle’s initial $\dot{x}$ velocity in $r_0/\mathrm{day}$ unit. The number of the test particles was $N=10^6$ in the case of Fig. \ref{Fig5} panel a and b, and $N=4 \times 10^6$ for panel c and d.
\par
In the first row of Fig. \ref{Fig5} the $(x, \dot{x}) \in [-1.014,-0.7142] \times [-0.2128, 0.21082]$ region was mapped. First the planar case of the R3BP was studied (Fig. \ref{Fig5} panel a). It is clearly visible that -- as expected -- the region around ${L_5}$ is stable (black zone) and there exits other stable regions as black area. The motion started exactly from ${L_5}$ is regular as it is well visible from Fig. \ref{Fig3}. The $x(t)$ of a test particle started exactly from the ${L_5}$ point (see the large cross in Fig. \ref{Fig5} panel a) is shown in Fig. \ref{Fig3}.
\par
The size of the quasi--periodic region next to point ${L_5}$ without the Sun (panel a) coincides with the stable region determined by a different method by \textit{\cite{Erdi2009}}. The largest distance of the initial points of the stable region from ${L_5}$ in the outward direction of the Earth is around 0.005 $r_0$ which is almost the same found by \textit{\cite{Erdi2009}}.
\par
In the next set of runs we have included the Sun and studied the orbits in the planar case. The most striking difference is the almost total disappearance of the stable region around ${L_5}$. The perturbation from the Sun destroys most of this stable zone and renders this region to a strongly chaotic area of the phase space. A small island remains on the left of the ${L_5}$. It is worth noting that other larger stable regions survived the perturbing effects of the Sun and remained stable with approximately the same structure and extension. A dashed line bounding box shows the region which was studied with more initial particles, that region is shown in panel c.
\par
In the second row of Fig. \ref{Fig5} we have explored the closer vicinity of ${L_5}$ with larger resolution (we have kept the ranges of the axes fixed to ease the comparison of the results). The reason to zoom in the phase space was to have a much finer resolution of the stable and chaotic regions in the vicinity of ${L_5}$. We have also increased the number of initial conditions to 4 million (the model for panel c was the same as that of panel b). In panel c a smaller plot shows a stable region shifted to the left of ${L_5}$. This region is stable for the integration time, but will vanish for longer times.
\par
In the last panel of Fig. \ref{Fig5} the computation was extended to the spatial case. One can see an extremely complex structure and also it useful to note that the escape rate is decreased with the introduction of the 3rd dimension. 

\par

\section{Chaos control}

Control of chaos refers to a process wherein a tiny perturbation is applied to a chaotic system, in order to obtain a desirable (chaotic, periodic, or stationary) behavior. The idea of chaos control was published at the beginning of this decade by \textit{\cite{Ott1990}}: the ideas for controlling chaos were outlined and a method for stabilizing an unstable periodic orbit was suggested, as a proof of principle. The basic idea consisted in waiting for a passage of the chaotic orbit close to the desired periodic behavior, and then applying a small fittingly chosen perturbation, in order to render an the chaotic motion more stable and predictable, which is often an advantage.
\par
According to the method developed by \textit{\cite{Ott1990}} (hereafter OGY method), first one obtains information about the chaotic system by analyzing a Poincar\`{e} section of the phase space. After the information about the section has been gathered, the system is integrated until it comes near a desired periodic orbit in the section. Next, the system is forced to remain on the target orbit by perturbing the appropriate parameter. One strength of this method is that it does not require a detailed model of the chaotic system but only some information about the Poincar\`{e} section. It is for this reason that the method has been so successful in controlling a wide variety of chaotic systems (e.g. turbulent fluids, oscillating chemical reactions, electronic circuits).
\par
There exits other alternative methods for chaos control. In general the strategies for the control of chaos can be classified into two main classes, namely: closed loop or feedback methods and open loop or non feedback methods. 
\par
The first class includes those methods which select the perturbation based upon a knowledge of the state of the system. Among them, we mention (besides \textit{\cite{Ott1990}}) the so called occasional proportional feedback simultaneously introduced by \textit{\cite{Hunt1991}} and \textit{\cite{Petrov1993}}, and the method introduced by \textit{\cite{Pyragas1992}}. All these methods are model independent, in the sense that the knowledge on the system necessary to select the perturbation can be done by simply observing the system for a suitable learning time.
\par
The second class includes those strategies which consider the effect of external perturbations (independent on the knowledge of the actual dynamical state) on the evolution of the system. Periodic (\cite{Lima1990}) or stochastic (\cite{Braiman1991}) perturbations have been seen to produce drastic changes in the dynamics of chaotic systems, leading eventually to the stabilization of some periodic behavior. These approaches, however, are in general limited by the fact that their action is not goal oriented, i.e. the final periodic state cannot be decided by the operator.

\subsection{Orbit--keeping procedure}

In our case the parameters of the system consist only the mass of the celestial bodies, therefore it is not possible to directly apply the OGY method to control the chaos. A different approach will be discussed and we stress that the method is based on observation and numerical experiments.
\par
Chaos is a double-edge sword, since it destabilizes the motion and renders it unpredictable, but at the same time it offers the opportunity to transport spacecrafts to distant locations with minimal fuel consumption. The present method is the result of an extensive numerical experiment. The idea was that if we cannot apply the OGY method, i.e. "kick" the body onto the stable manifold of the hyperbolic point, then let us try to drive it away from that point. In order to accomplish this a small force (thrust) was added to the equation of motion of the test particle:
\begin{equation*}
|\vec{f}| = \vec{e}_{\mathrm v} \left\{ \begin{array}{ll}
         F_0 & \mbox{if $t \in \left[t_{ci},\, t_{ci} + \Delta t_i \right]$}, \\
         0   & \mbox{otherwise},\end{array} \right., \quad i = 1,\ldots
\end{equation*}
where $F_0 > 0$ is the magnitude of the applied force and $\vec{e}_{\mathrm v}$ is the unit vector in the direction of the particle's velocity, $t_{ci}$ is the time instant when the control begins and $\Delta t_i$ is the length of the control.
\par
\begin{figure}
\includegraphics[width=\linewidth]{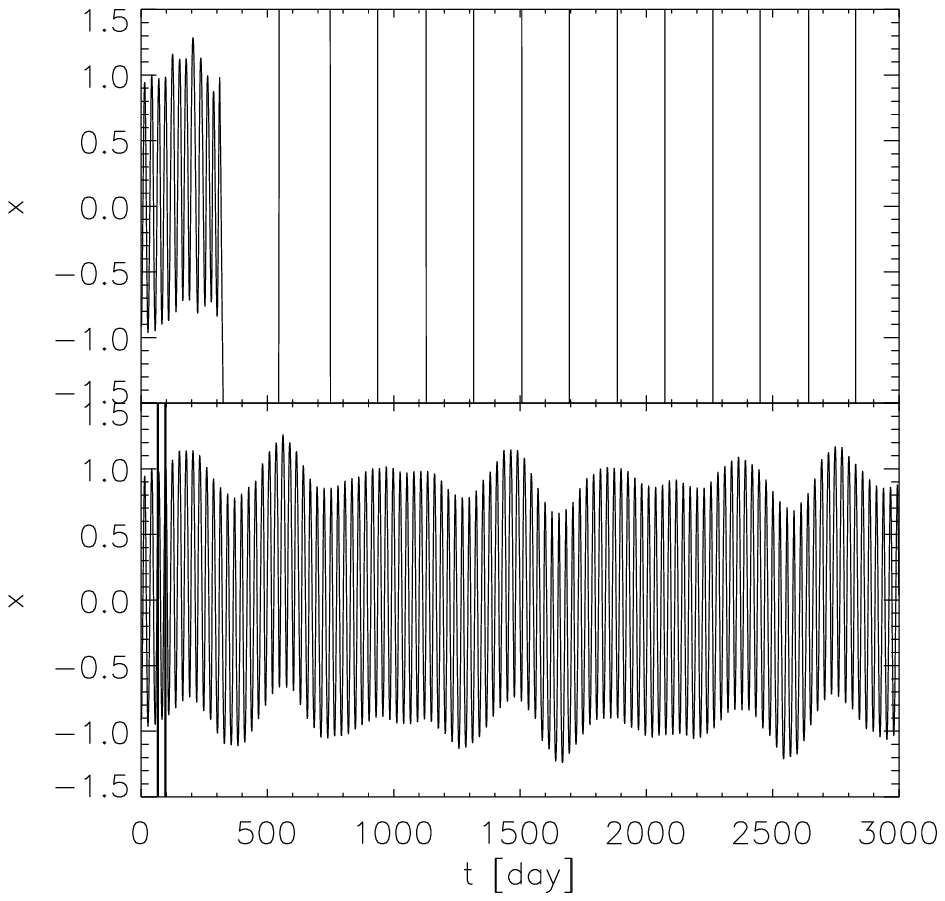}
\caption{The $x(t)$ of the probe in the R4BP without (upper panel) and with control (lower panel). The probe was started from the vicinity of the ${L_5}$ point (see Table \ref{Table1}). The orbit without control stays around ${L_5}$ for approximately 300 days, then it leaves that region. We have applied a thrust from $t_{c1} = 66.32$ days for $\Delta t_1 = 29.87$ days. This control stabilizes the motion up to approximately 5000 days. On the lower panel the vertical lines denote the start and stop of the control.} 
\label{Fig6}
\end{figure}

As a proof of our concept we will show a test case where we have applied our orbit--keeping procedure for a spacecraft to keep it around the ${L_5}$ point. In Fig. \ref{Fig6} we have plotted the time evolution of the $x$ coordinate of the spacecraft (its initial conditions are in the last column of Table \ref{Table1}). In the upper panel the probe orbits without control and stays around ${L_5}$ for approximately 300 days, then it escapes that region and begins to oscillate widely. In the lower panel the $x(t)$ is plotted when the above defined force (thrust) was applied from $t_{c1} = 66.32$  days for $\Delta t_i = 29.87$ days (see the vertical lines), which corresponds to about one period of the Moon. The value of $F_0 = 9.01 \times 10^{-5}$ was used in this case. This value is the result of dozens of tests and proved to be a quite general value and not only the "best" for the present situation. We have also investigated braking force ($F_0 < 0$) but this alternative in our tests never stabilized the orbit.
\par
\begin{figure}
\includegraphics[width=\linewidth]{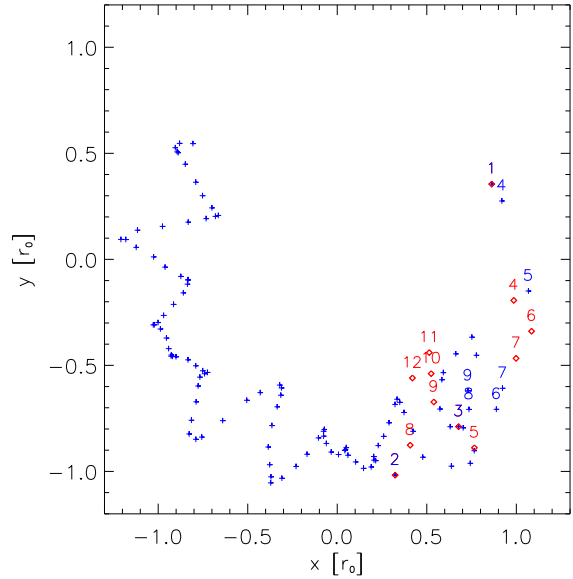}
\caption{The $x-y$ section with $z=0$ and $\dot{z} > 0$ of the test particle’s orbit from $t=0$ to $t \leq 3000$ days. The red diamond symbols are the section points of the trajectory without control. The blue plus sign symbols are the section points of the trajectory with control. The numbers over the symbols are the $n$th section point. No 1, 2 and 3 points are overlapping, since the control starts at point No. 3. The Earth is in the centre.} 
\label{Fig7}
\end{figure}
\begin{figure}
\includegraphics[width=\linewidth]{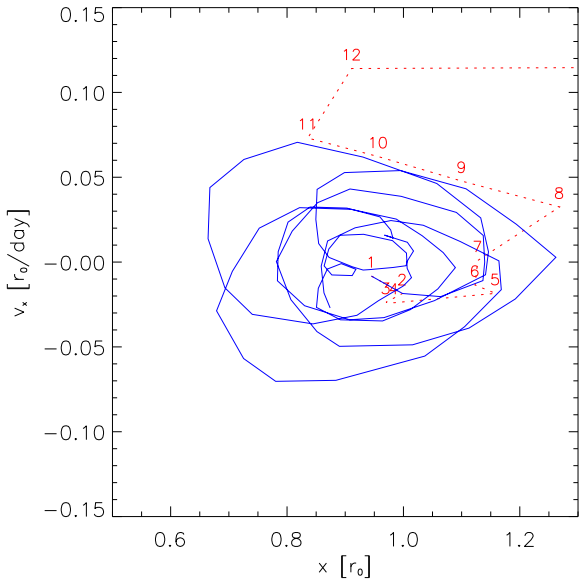}
\caption{The $x-\dot{x}$ section with $y=0$ and $\dot{y} > 0$ of the test particle’s orbit from $t=0$ to $t \leq 3000$ days. The red dotted line connects the section points of trajectory without control, while the blue continuous line connects those with control. The numbers are the $n$th section point of the trajectory without control. No 1 and 2 are overlapping, since the control starts at point No. 3.} 
\label{Fig8}
\end{figure}

To determine the $i$th start time of the control $t_{ci}$ one has to integrate the motion of the probe and monitor its orbit. The $x(t)$ function and the $x-y$ and $x-\dot{x}$ sections are proved to be useful in order to determine the start and stop time of the control. Our numerical experience has shown us that the control must start well before the escape and at a moment when the orbit intersects the $x-y$ plane, i.e. when the $z = 0$ and $\dot{z} > 0$ conditions are met. To find the "best" starting time is the result of a systematic search but the structure or pattern of the points on the $x-y$ and $x-\dot{x}$ section provides some additional information. In Fig. \ref{Fig7} the $x-y$ section with $z=0$ and $\dot{z} > 0$ of the test particle’s trajectory from $t=0$ to $t \leq 3000$ days is shown. The orbit without control is represented by red diamonds while the one with control is shown by blue plus signs. The first few section points are numbered to see the time evolution of them. In Fig. \ref{Fig8} the $x-\dot{x}$ section with $y=0$ and $\dot{y} > 0$ of the test particle’s trajectory from $t=0$ to $t \leq 3000$ days is demonstrated. The blue continuous line segments connects the successive section points of the trajectory with control, the red dotted line segments connects the points without control. The controlled trajectory's section points stay inside a bounded region whereas those without control leave this region after a few period of the Moon.
\par
In the present example shortly after the first section point the sequence of points do not follow any "regular" pattern, (i.e. a "scalloped edge" ellipse or a snowflake on the $x-y$ section plane, see the blue plus signs in Fig. \ref{Fig7} or a distorted ellipse on the $x-\dot{x}$ section plane, see the blue continuous line in Fig. \ref{Fig8}) therefore we switch on the additional force immediately at point No. 3. There are two other parameters which we have to set: ({\it i}) the force $F_0$ and ({\it ii}) the length of the control $\Delta t_i$.
\par
We have experiment with the magnitude of the force $F_0$ and found that the value of $F_0 \approx 9.01 \times 10^{-5}$ could be applied in all the cases we have tried.
\par
In all our cases the start and stop time of the control is a time instant of a section in the $x-y$ section plane. In our test cases the stop time was either the next, the second or the third section after the start time, so the duration of the control $\Delta t_i$ is equal to the time span between two or more subsequent points on the $x-y$ section plane. The time span between two successive section point is approximately equal to the period of the Moon, therefore loosely speaking we may measure the length of the control in "periods". One period of a control means the time difference between two successive section points. We emphasize that these results based solely on numerical observations and experimentation: we have extensively studied other $\Delta t_i$s but it turned out that the best choice was always one of the moment of a successive section.
\par
\begin{figure}
\includegraphics[width=\linewidth]{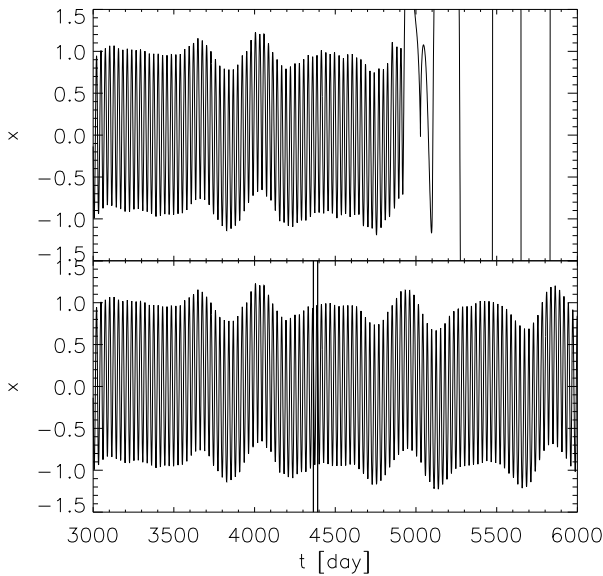}
\caption{The $x(t)$ of the test particle. The orbit after the first control without the next control stays around ${L_5}$ for approximately 5000 days, then it leaves that region. In the lower panel we have applied a thrust from $t_{c2} = 4364.28$ days for $\Delta t_2 = 26.6$ days. This control stabilizes the motion up to approximately 18 900 days. The thick vertical lines denote the start and stop of the control.} 
\label{Fig9}
\end{figure}

The control in the test case was started at the 3rd point on the $x-y$ section plane (see Fig. \ref{Fig7}) which corresponds to time $t_{c1} = 66.32$ days and the thrust was applied until the next section was detected, i.e. for one period. Having switched off the thrust we have integrated the system and monitored the orbit of the spacecraft. As it can be seen form Fig. \ref{Fig9} the $x(t)$ function oscillates quasi periodically about 0 and its amplitude does not varies significantly ($x(t)$ in Fig. \ref{Fig9} is a continuation of $x(t)$ in Fig. \ref{Fig6}). This behavior can be observed till $t \approx 5000$ days when the probe leaves the vicinity of ${L_5}$ again. Just as in the first case, we have applied a thrust from $t_{c2} = 4364.28$ days until we have detected the next section, which resulted in $\Delta t_2 = 26.6$ days (somewhat shorter than the in previous control). The same value of $F_0$ was used again.
\par
\begin{figure}
\includegraphics[width=\linewidth]{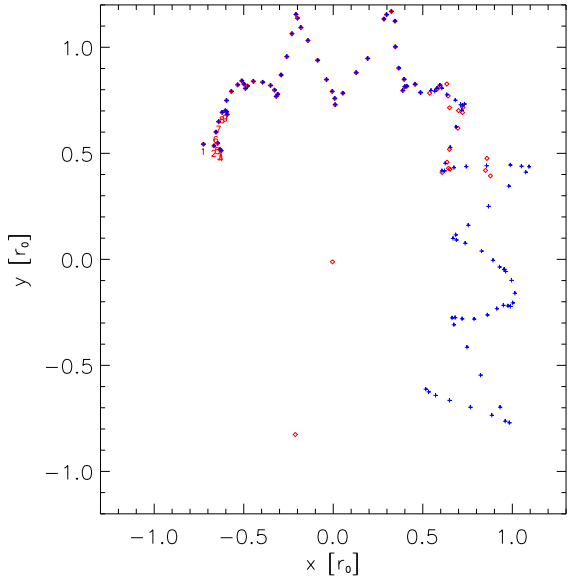}
\caption{The $x-y$ section with $z=0$ and $\dot{z} > 0$ of the test particle’s orbit from $t=3000$ days to $t \leq 6000$ days. The symbols are the same as in Fig. \ref{Fig7}. The Earth is in the centre.} 
\label{Fig10}
\end{figure}
\begin{figure}
\includegraphics[width=\linewidth]{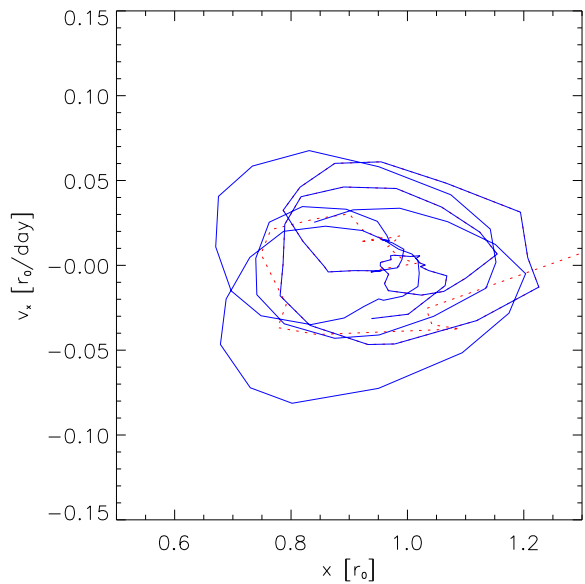}
\caption{The $x-\dot{x}$ section with $y=0$ and $\dot{y} > 0$ of the test particle’s orbit from $t=3000$ days to $t \leq 6000$ days. The symbols are the same as in Fig. \ref{Fig8}.} 
\label{Fig11}
\end{figure}

The trajectory on the $x-y$ and $x-\dot{x}$ section planes are shown in Fig. \ref{Fig10} and Fig. \ref{Fig11}, respectively. In Fig. \ref{Fig10} we see the continuation of the pattern that was observed in Fig. \ref{Fig7} and in Fig. \ref{Fig11} we see the continuation of Fig. \ref{Fig8}. In Fig. \ref{Fig10} the $x-y$ section with $z=0$ and $\dot{z} > 0$ of the test particle’s trajectory from $t=3000$ days to $t \leq 6000$ days is shown, the symbols are the same as in Fig. \ref{Fig7}. The red diamonds and the blue plus signs overlap until when the control was switched on and then the points depart from each other. The start time and length of the control is again a result of a systematic search: $t_{c2}$ and $\Delta t_2$ were found to result in a quasi periodic orbit for $t$ = 18 900 days. In Fig. \ref{Fig11} the $x-\dot{x}$ section with $y=0$ and $\dot{y} > 0$ of the test particle’s trajectory from $t=3000$ days to $t \leq 6000$ days is shown. The controlled trajectory's section points (blue continuous line) stay inside a bounded region whereas those without control (red dotted line) leave this region after a few period of the Moon. We note that the size of the bounded region hardly changed comparing it with the previous time span shown in Fig. \ref{Fig8}. The result of the systematic search was that the start time of the second control was $t_{c2} = 4364.28$ days and the length of the control was one period ($\Delta t_2 = 26.6$ days).
\par
We used a third and a fourth control beginning at $t$ = 18899.49 days and $t$= 18979.37 days which resulted in bounded motion until $t$ = 30196.72 days. We note that the duration of the third control was again one period but three periods for the fourth control. The results in terms of the $x(t)$, $x-y$ and $x-\dot{x}$ are very similar to the ones presented above, so we do not show them.
\par
The orbit--keeping procedure can be summarized as follows:
\begin{enumerate}
 \item Determine the moment of "escape": the motion of the probe is integrated until it leaves the region allowed by some prescription (i.e. the mission instructions).
 \item A systematic search is carried out in order to find the best or acceptable time instant $t_{ci}$ when the control should start. According to our numerical observations ({\it i}) this time instant should correspond to the moment of a section in the $x-y$ plane and ({\it ii}) it must precede with a few to a few dozens of section points the time of escape.
 \item Another systematic search is carried out in order to find the length of the control $\Delta t_i$: according to our observation the stop of the control is also determined by the moment of section on the $x-y$ plane; in our cases this section point was the first, the second or the third after the start time.
 \item We have found that $F_0 \approx 9 \times 10^{-5}$ was applicable in all our test cases.
\end{enumerate}
\par
This process was tried for several initial conditions and we could stabilize the motion of the spacecraft in all test cases: the probe was temporary pushed on a bounded orbit. In the demonstrated case the time intervals of the quasi periodic  motion were long enough for a usual space mission to carry out. After the first control it was approximately 13 years, then 38 years and after the last control 31 years. We have repeated our test cases with probes having different masses: the start time of the control and its duration (29.87+26.6 days that resulted quasi periodic motion for 13 + 38 years) was the same for probes as massive as $10^{6}$ kg. This is understandable since the most massive probe's influence on the dynamics of the other bodies of the model is still negligible. We have estimated the energy needed for the control: for a one period length control of 1 kg mass satellite the consumed energy is approximately 12 kJ.

\section{Conclusion}

We have applied the concept of SMD and the basic idea of the OGY method to stabilize the motion of a spacecraft around the ${L_5}$ in the R4BP. The phase space structure in the vicinity of ${L_5}$ was mapped in both the R3PB and in the R4BP. We have studied the escape rate of the trajectories and fitting a line to the measured data we were able to estimate the average lifetime of the particles which turned out to be 446 (R3BP) and 307 (R4BP) days. Integrating the two different models we mapped the phase space around the ${L_5}$ point of the Earth--Moon system.
\par
Using the phase space portrait our main goal was to apply the OGY method to keep a spacecraft close to the vicinity of ${L_5}$. The OGY method cannot be directly applied to our problem at hand (there are no variable system parameters) therefore we modified the equation of motions with the addition of a time dependent force to the motion of the spacecraft. There are three free parameters of our orbit--keeping procedure: ({\it i}) the magnitude of the thrust ($F_0$), ({\it ii}) the start time of the control and ({\it iii}) the length of the control. We have performed extensive numerical experiments and according to our observations we found the following rule of thumb useful: ({\it i}) $F_0 \approx 9 \times 10^{-5}$ is applicable in all cases, ({\it ii}) the start time of the control is a time instant of a section in the $x-y$ section plane and ({\it iii}) the length of the control is an integer multiply of the period (the time between two successive section points).
\par
We note that we lack the physical background and explanation of our orbit--keeping procedure. We are fully aware of the fact that the underlying physics and the properties of the chaos responsible for the observed behavior must be carefully studied in the future. We are performing systematic research in this direction in order to understand the basic processes causing the observed behavior.

\acknowledgements
TK has been supported by the Lend\"ulet-2009 Young Researchers' Program of the Hungarian Academy Sciences and the ESA PECS Contract No. 4000110889/14/NL/NDe. \'AS is grateful to the OTKA-103244 support. The authors want to thank B. \'Erdi and T. T\'el for the helpful discussions.


\begin{thebibliography}{aaa}

\bibitem[{Braiman \& Goldhirsch 1991}]{Braiman1991} Y. Braiman, J. Goldhirsch (1991), Taming chaotic dynamics with weak periodic perturbations, Phys. Rev. Lett. 66, 2545

\bibitem[{Danby 1964}]{Danby1964} J.M.A. Danby (1964), Stability of the triangular points in the elliptic restricted problem of three bodies. Astron. J. 69, 165–172 

\bibitem[{Dormand \& Prince 1978}]{Dormand1978} J.R. Dormand, P.J. Prince (1978),  New Runge-Kutta algorithms for numerical simulation  in dynamical astronomy Celestial Mechanics 18, 223

\bibitem[{Dunham \& Farquhar 2003}]{Dunham2003} D. W. Dunham,  R. W. Farquhar (2003), Libration Point Missions, 1978-2002. In Libration Point Orbits and Applications, edited by G. G\'omez, M. W. Lo, J. J. Masdemont, World Scientific. pp: 48-73. 

\bibitem[{\'Erdi et al. 2009}]{Erdi2009} B. \'Erdi, E. Forg\'acs-Dajka, I.Nagy, R. Rajnai (2009), A parametric  study of stability and resonances around ${L_5}$ in the elliptic restricted three-body problem Celestial Mechanics and Dynamical Astronomy,  104, 145 

\bibitem[{de Fillipi 1978}]{Fillipi1978} G. de Fillipi (1978), Station Keeping at the L4 Libration Point: A Three Dimensional Study. Master’s thesis, Dept. Aeronautics and Astronautics, Air Force Institute of Technology Wright-Patterson AFB, Ohio, USA 

\bibitem[{G\'omez et al. 2001}]{Gomez2001} G. G\'omez, J. Llibre, R.  Mart\'{\i}nez  and C. Sim\'o (2001), Dynamics and Mission Design near Libration Points - Volume 1. Fundamentals: The Case of Collinear Libration Points. World Scientific, Singapore

\bibitem[{Howell et al. 1998}]{Howell1998} K. C. Howell, B. T. Barden,  R. S. Wilson,  M. W. Lo (1998), Trajectory Design Using a Dynamical Systems Approach with Application to GENESIS. Advances in the Astronautical Sciences 97: 1665-1684

\bibitem[{Hunt 1991}]{Hunt1991} E. R. Hunt (1991), Stabilizing High-Period Orbits in a Chaotic System: the Diode Resonator, Phys. Rev. Lett. 67, 1953

\bibitem[{Kov\'acs \& \'Erdi 2009}]{Kovacs2009} T. Kov\'acs, B. \'Erdi (2009), Transient chaos in the Sitnikov problem, Celestial Mechanics and Dynamical Astronomy, 105, 289  

\bibitem[{Lai \& T\'el 2011}]{Lai2011} Y.C. Lai, T. T\'el (2011), Transient Chaos, Springer

\bibitem[{Lima \& Pettini 1990}]{Lima1990} R. Lima, M. Pettini (1990), Experimental suppression of chaos in a modulated multimode laser, Phys. Rev. A 41, 726

\bibitem[{Murray \& Dermott 1999}]{Murray1999} C. D. Murray, S. f. Dermott  (1999), Solar System Dynamics, Cambridge University Press  

\bibitem[{O'Neill \& Gerard 1974}]{Neill1974} O'Neill, K. Gerard (1974), The Colonization of Space,  Physics Today 27 (9): 32–40

\bibitem[{Ott, Grebogi \& Yorke 1990}]{Ott1990} E. Ott, C. Grebogi, J. A. Yorke (1990), Controlling Chaos, Phys. Rev. Lett. 64, 1196 
 
\bibitem[{Perozzi \& Ferraz-Mello 2010}]{Perozzi2010} E. Perozzi, S. Ferraz-Mello (2010), Space Manifold Dynamics, Springer 

\bibitem[{Petrov et al. 1993}]{Petrov1993} V. Petrov, V. G\'aspar, J. Masere, K. Showalter (1993), Controlling chaos in the Belousov—Zhabotinsky reaction,  Nature 361, 240

\bibitem[{Pyragas 1992}]{Pyragas1992} K. Pyragas (1992), Control of Chaos via an Unstable Delayed Feedback Controller, Phys. Lett. A 170, 421
 
\bibitem[{Salazar et al. 2012}]{Salazar2012} F.J.T. Salazar, C. F. de Melo, E. E. N. Macau, O. C. Winter (2012), Three-body problem, its Lagrangian points and how to exploit them using an alternative transfer to {$ L_4$} and ${L_5}$, Celest. Mech. Dyn. Astr. 

\bibitem[{Schutz 1977}]{Schutz1977} B. F. Schutz (1977),  Orbital mechanics of space colonies at ${L_4}$ and ${L_5}$ of the earth-moon system. In: Aerospace Sciences Meeting 15., 1977, Los Angeles, California. Paper No. 77-33, Reston, Virginia: American Institute of Aeronautics and Astronautics, 1977. p. 12. 3

\bibitem[{Szebehely 1967}]{Szebehely1967} V. Szebehely (1967), Theory of Orbits, Academic press, New York    

\bibitem[{T\'el \& Gruiz 2006}]{Tel_2006} T. T\'el, M. Gruiz (2006), Chaotic Dynamics, Cambridge University Press 


\end{thebibliography}
\end{document}